# The impact of a few: The effect of alternative formulas for recruiting talent in a non-competitive system[1]

Nicolas Robinson-Garcia[*], Evaristo Jiménez-Contreras[*] and Clara Calero-Medina[**]

[*]*elrobin@ugr.es; evaristo@ugr.es*
EC3 Evaluación de la Ciencia y de la Documentación Científica, Universidad de Granada, Colegio Máximo de Cartuja, s/n, Granada, 18071 (Spain)

[**]*clara@cwts.leidenuniv.nl*
Centre for Science and Technology Studies, Leiden University (The Netherlands)

## *Introduction*

The introduction of performance-based university research funding systems (hereafter PRFS) has widely spread and established in many countries since the 1980s. Such systems have been developed in order to allocate funds at the institutional level, leading to differences between academic institutions and a concentration of talent in those which rank higher in this evaluation benchmarks. Good examples are the Research Evaluation Framework (former Research Assessment Exercises) in the United Kingdom (Martin, 2011), or the Australian Excellence in Research Initiative (Butler, 2003). A recent review on the characteristics of PRFS and their advantages and limitations is offered by Hicks (2012). But other countries have adopted another approach, focusing on individual incentives rather than institutional evaluation. The reason for doing so is that differences between universities are not as significant as differences between individuals within universities (Abramo, Cicero & D'Angelo, 2011).

This is the case of Spain, which has implemented an evaluation exercise by which the research performance of tenured staff is periodically assessed (Jiménez-Contreras, Moya-Anegón & Delgado López-Cózar, 2003). However, mobility within universities is seriously constrained, showing high rates of inbreeding due to their restrictive employment conditions (Navarro & Rivero, 2001). Such conditions discourage foreign researchers (Pickin, 2001) as well as preventing Spanish researchers working abroad from returning to their home country. As a response to this situation, some programs have been developed at the national and regional level. The most studied one is the Ramón y Cajal national program developed in 2001 which aims at providing the means to cease with what was considered as a 'brain drain' of young Spanish researchers. But the outcome of the Ramón y Cajal program has left mixed feelings (Mandavilli, 2006; Sanz-Menéndez, 2007), being considered as a missed opportunity to overcome the rigidities of the traditional hiring system.

However, little is known about the effects of similar regional formulas. This is the case of the *Institució Catalana de Recerça i Estudis Avançats* (henceforth ICREA) created in 2001 by the Catalan regional government. ICREA is a recruiting agency focused on attracting senior researchers with an international research background. One can hypothesize on the relevance of such institution when analyzing the leading role Catalan universities play in the national system (Torres-Salinas et al., 2011). In this paper we intend to explore the influence exerted by researchers hired through ICREA (henceforth ICREA researchers) on Catalan universities.

---

[1] Nicolás Robinson-García is currently supported by a FPU (Formación de Profesorado Universitario) grant of the Spanish Ministerio de Economía y Competitividad.

The functioning of this agency can be resumed as follows. ICREA establishes collaboration agreements with any research center or university within Catalonia willing to host ICREA researchers, covering their salary on the condition that the host institution will provide them the needed infrastructure and integrate them within the institution. Hence, every year ICREA offers a closed number of positions for which candidates must have previously come in terms with their potential host institution. Then, these are evaluated through a demanding peer review process undertaken by five panels each of them corresponding to a scientific field, and the ones with the highest ratings, despite the selection panel or the host institution, are then selected. On their behalf, ICREA researchers are expected to create new research lines, develop and lead their own research groups and integrate themselves within their host institution. Figure 1 briefly resumes the selection process of ICREA researchers.

Figure 1. Flowchart of the selection process of ICREA researchers

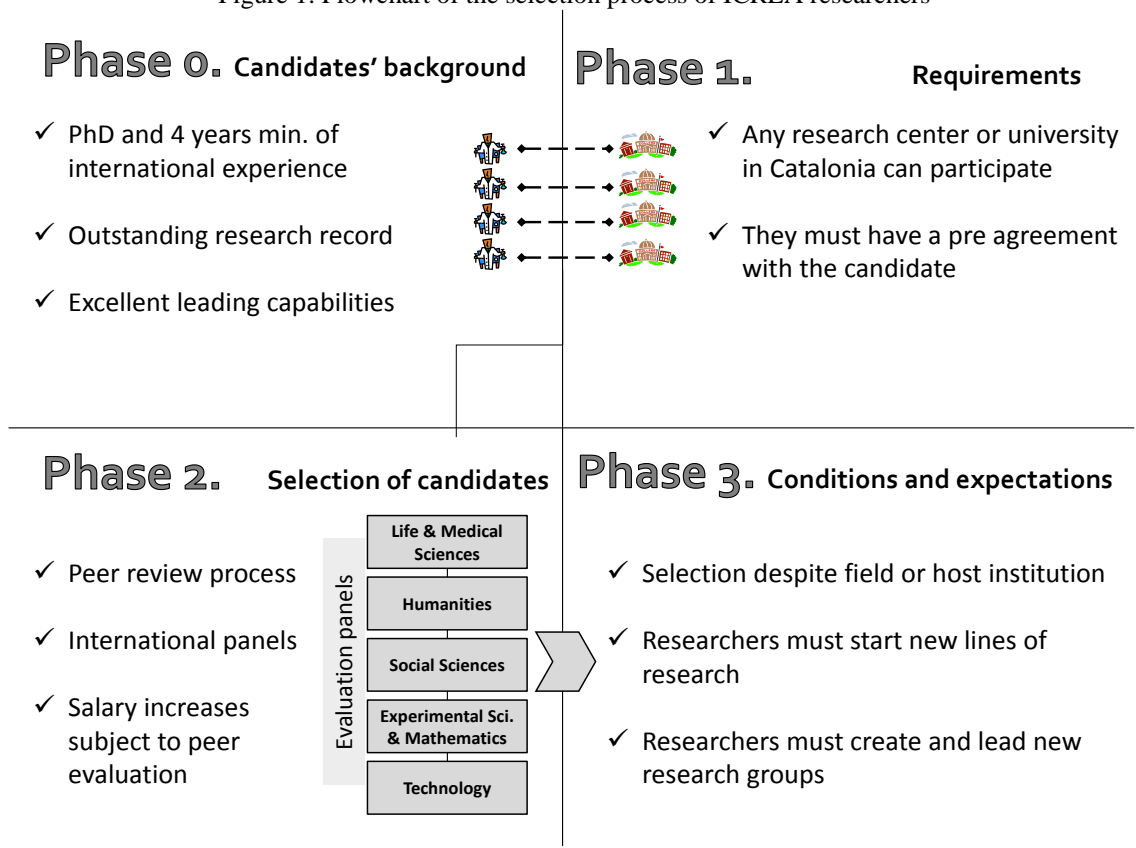

Since its set up, ICREA has hired up to 294 researchers (270 currently active according to the Catalan Research and Innovation Plan 2010-2013) and it is highly considered by both, ICREA researchers and their host institutions (Technopolis, 2011). It nourishes three types of institutions: research centres belonging to the National Research Council (CSIC), public universities and research centres belonging to the Catalan Government (known by the acronym CERCA) which were develop at the same time as ICREA. In this study we focus on the strategies followed by universities to allocate these researchers and their impact in specific disciplines. Our research questions are:

- Can we perceive through the number of publications strategic policies when allocating ICREA researchers by universities? Are they reinforcing specific departments or fields?

- Are these researchers making a difference in terms of highly cited papers? How much do they contribute within their field and within their university?

## *Material and methods*

In this paper we analyze the contribution of ICREA researchers to Catalan universities within the 2002-2012 time period. A *top-down* approach is used, based on the dataset of publications from the Leiden Ranking (Waltman et al., 2012) included in the CWTS in-house version of the Web of Science database. ICREA researchers are affiliated to both, their host institution and ICREA. Hence, to identify ICREA publications, we searched in the address field to all possible variants ICREA and identified a subset of publications linked to both, ICREA and Catalan universities . This subset was later checked with the information shown in the annual reports available in the website of ICREA since 2008 (http://www.icrea.cat/). We obtained 84.69% of the total output reported by the agency, although these reports display the number of publications for each of the five fields in which the selection panels are divided and do not report of possible overlaps, document types included or if they include other citation indexes such as the Conference Proceedings Citation Indexes.

In order to show a general overview, we used the five areas defined in the latest edition of the Leiden Ranking. Then, to locate detailed areas (subjects categories) with a higher concentration of ICREA researchers we focused on all subject categories for each university with at least 50 publications (articles and reviews) for the study time period, and selected those where ICREA publications represented at least 20% of the total share. For this subset, we identified the share of publications within the top highly cited papers in each field. Here we define highly cited papers as the publications cited equal or more than the 90th percentile limit of the citation distribution (top 10%).

## *Results*

Between 2002 and 2012, the Catalan universities produced a total of 77,547 publications. ICREA publications represent 4,6% of the total share. Barcelona is the university with the largest production (32,233 publications), followed by Autónoma de Barcelona (24,975). The third most productive university is Rovira i Virgili (5,781). On the other end we find Lleida (2,684). The university with a higher representation of ICREA publications is Pompeu Fabra, where these account for 13.7% of its total output. Next, we have Autónoma de Barcelona (5.9%) followed by Rovira i Virgili (4.4%).

Table 1. Output of Catalan universities and distribution of ICREA publications for the 2002-2012 time period

| University | Biomedical & Health Sci | | Life & Earth Sci | | Maths & Computer Sci | | Natural Sci & Engineering | | Social Sci & Hum | |
|---|---|---|---|---|---|---|---|---|---|---|
| | Pub | %ICREA | Pub | %ICREA | Pub | %ICREA | Pub | %ICREA | Pub | %ICREA |
| Barcelona | 16350 | 1.72 | 5840 | 2.35 | 1332 | 4.05 | 9513 | 6.83 | 2538 | 3.23 |
| Autónoma Barcelona | 12385 | 1.89 | 5368 | 4.79 | 2114 | 5.44 | 5588 | **15.44** | 2283 | 5.43 |
| Politecnica Catalunya | 792 | 2.53 | 1719 | 0.93 | 3851 | 0.91 | 4912 | 2.75 | 298 | 1.68 |
| Rovira i Virgili | 1706 | 1.88 | 1265 | 7.75 | 833 | 0.36 | 2419 | 6.12 | 539 | 2.97 |
| Pompeu Fabra | 2720 | **13.42** | 630 | **19.52** | 707 | **13.15** | 291 | 14.09 | 1228 | **9.20** |
| Girona | 968 | 1.14 | 1220 | 2.05 | 536 | 1.12 | 1470 | 7.28 | 389 | 0.00 |
| Lleida | 850 | 2.00 | 1356 | 3.76 | 262 | 0.00 | 477 | 0.00 | 209 | 0.48 |

Note: Higher shares of ICREA output are highlighted in bold.

However, if we observe how these publications are distributed by broad areas, we observe significant differences. In table 1 we show the total output by university and area, along with the distribution by areas of ICREA publications. Here the percentage of ICREA output is again higher for Pompeu Fabra for all areas with the exception of Natural Sciences & Engineering, where a 15.4% of publications from Autónoma de Barcelona belong to ICREA researchers, followed by Pompeu Fabra (14.1%).

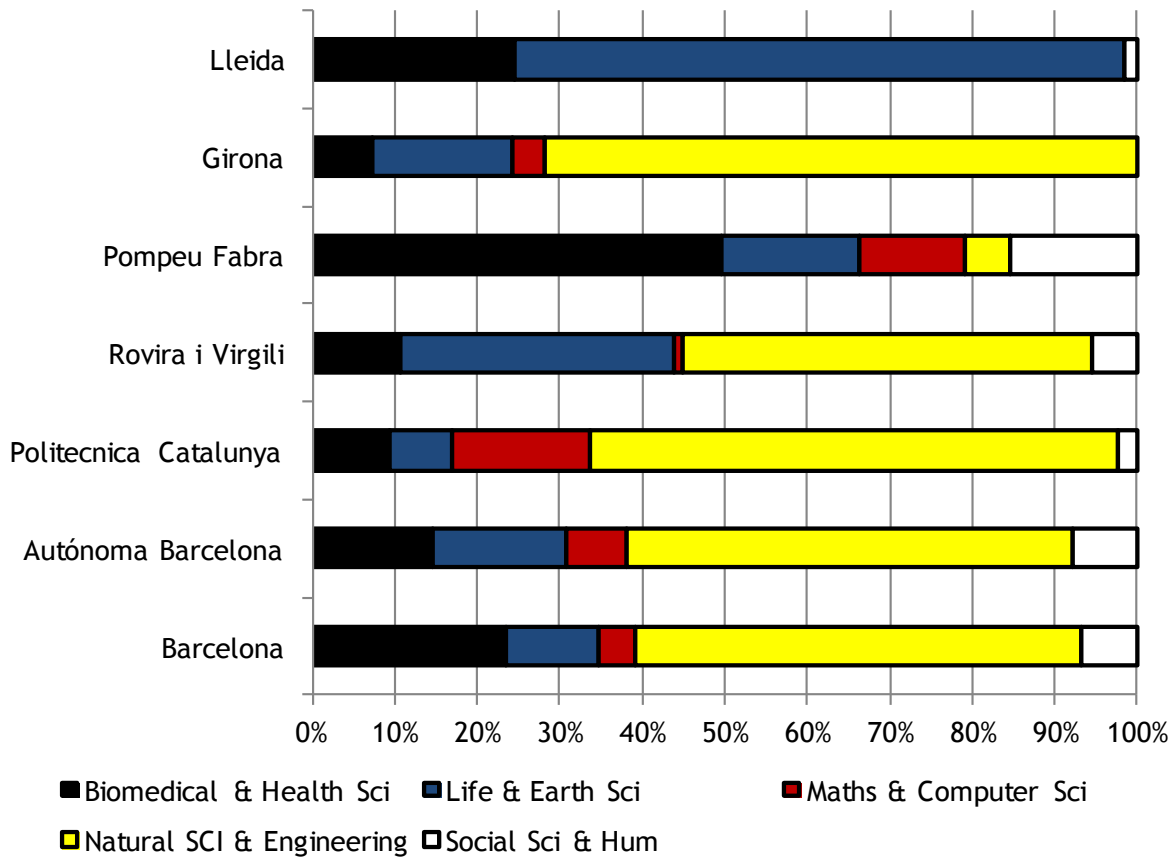

Figure 2. Distribution of ICREA publications by broad fields

As observed in figure 2, ICREA publications are distributed similarly in most universities. Natural Sciences and Engineering is the area where most publications are concentrated (more than 50%), and then the second area varies according to the university. The only exceptions are Pompeu Fabra and Lleida. The former allocates 54.5% of ICREA publications within the area of Biomedical & Health Sciences, while the latter concentrates 86.4% of its publications in Life & Earth Sciences.

If we focus on subject categories in order to find those where ICREA publications represent a significant share of the total output, we find that only four universities (Autónoma de Barcelona, Girona, Pompeu Fabra and Rovira i Virgili) meet the criteria for inclusion. In figure 3 we show the subject categories by university with at least 50 publications during the study time period in which ICREA publications represent at least 20% of the total share. Pompeu Fabra is the university with a larger number of subject categories where ICREA publications play an important role on the overall performance. As observed, 12 subject categories, almost all of them related with Biomedicine and Health Sciences, meet the inclusion criteria. The next university with a higher number of subject categories is Autónoma de Barcelona with 9 subject categories. Here, most categories relate with Physics.

Figure 3. Share of ICREA publications by university and by subject categories for the 2002-2012 period. Only subject categories with ≥ 50 publications and ≥ 20% of ICREA output are shown

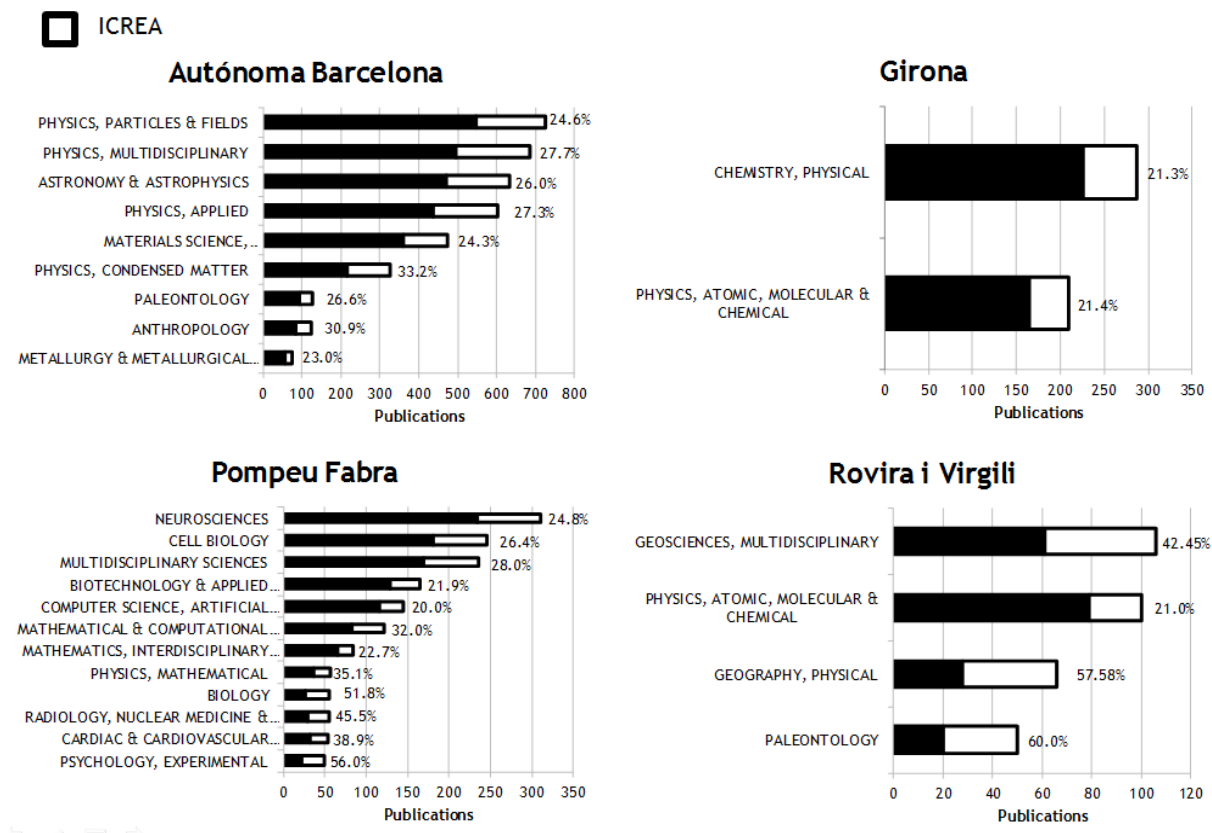

In figure 4 we look at the share of highly cited papers for ICREA publications, not ICREA publications and the overall for the universities and subject categories shown in figure 3. As observed, with the exception of Rovira i Virgili, in the rest of the cases, ICREA publications exceed the rest of the publications, having a positive impact on the overall value of the indicator. The only exception where ICREA publications do not perform better than the rest are Anthropology in Autónoma de Barcelona, where there is almost no difference between both subsets of publications and all subject categories with the exception of Paleontology in Rovira i Virgili. Here, not ICREA output exceeds ICREA output according to the % of highly cited papers.

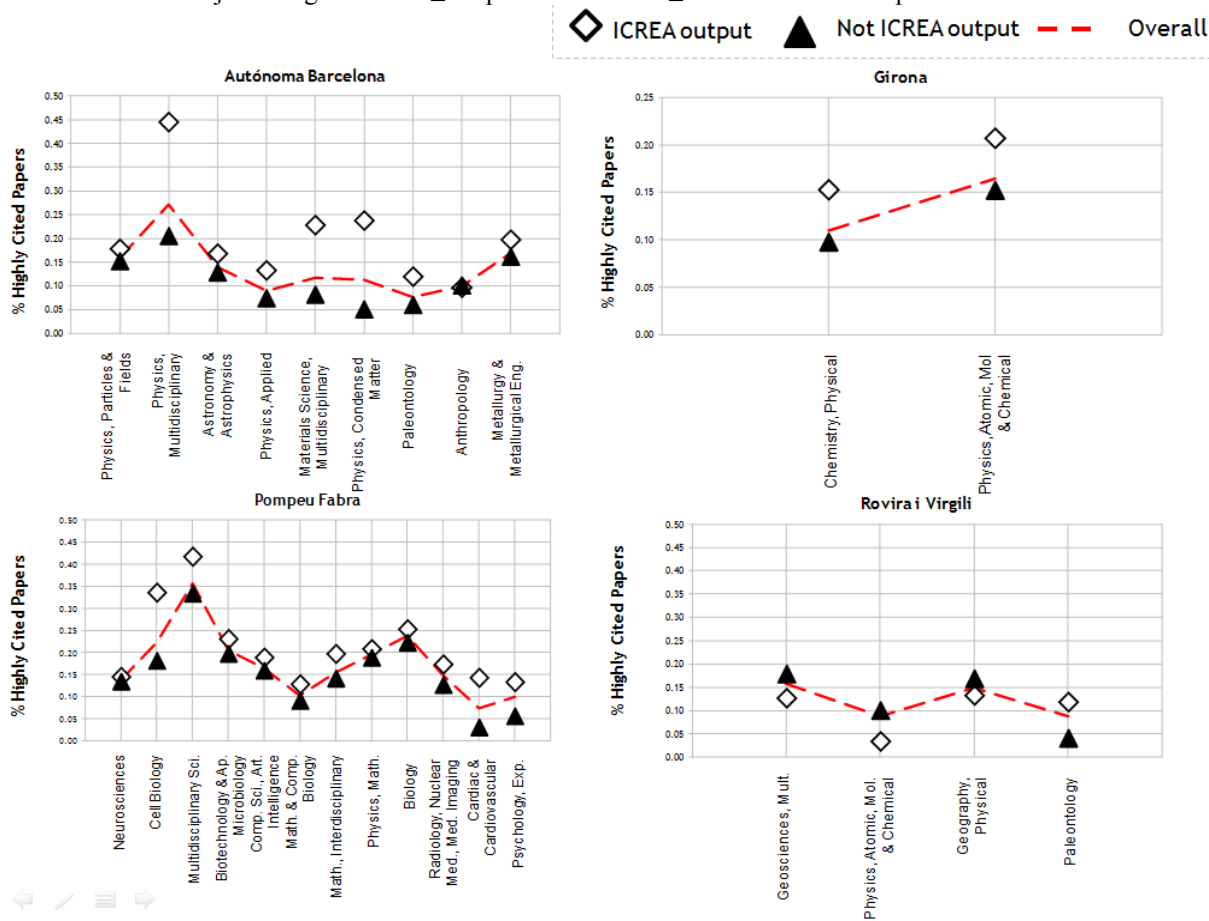

Figure 4. Percentage of highly cited papers by university and subject categories for the 2002-2012 period. Only subject categories with ≥ 50 publications and ≥ 20% of ICREA output are shown

## Discussion and concluding remarks

In this paper we conduct an exploratory analysis on strategies adopted by Catalan universities to recruit ICREA researchers. For this, we identified all ICREA publications for the 2002-2012 time period. Our results show that not all universities are using this formula to strategically reinforce the research performance of specific disciplines. Formulas such as ICREA allow Spanish universities, which present low levels of governance (Aghion, et al., 2010), a chance to develop research policy strategies to reinforce certain departments and scientific fields. In this sense, four out of seven Catalan universities seem to have included systematically ICREA researchers specialized in certain fields (Autónoma de Barcelona, Pompeu Fabra, Rovira i Virgili and Girona). The most clear pattern is that shown by Pompeu Fabra which shows a special interest on hiring ICREA researchers in the Biomedical Sciences. This may be due to their specialized profile in this area and may partly explain the unexpected growth in terms of research impact of this university (Robinson-García et al., 2013). Such results indicate a perceived recruitment policy towards hiring ICREA researchers in these universities.

Regarding the impact of ICREA researchers on the performance of universities in these strategic areas, as expected, we observe that in most cases, ICREA output shows higher values when focusing on the share of highly cited papers. The only exception is Rovira i Virgili, a surprising that should be better analyzed in order to understand what exactly is happening there. Here we have used the highly cited papers indicator to determine research excellence, however, other bibliometric indicators should be analyzed to determine the level of success of the ICREA initiative on the enhancement of research performance. Also, as a *top-down* approach was taken, much information which would allow us to determine issues such when did researchers join each university is missing. Further research should include bio data for ICREA researchers, in order to further understand the implications of such policy within the regional university system. Also, it would be desirable to analyze and compare ICREA with other similar recruiting formulas implemented in other regions as well as with the national Ramón y Cajal program.